# Alloying Effects on the Microstructure and Properties of Laser Additively Manufactured Tungsten Materials

W. Streit Cunningham[1], Eric Lang[2], David Sprouster[1], Nicholas Olynik[1], Ajith Pattammattel[3], Daniel Olds[3], Khalid Hattar[4], Ian McCue[5], and Jason R. Trelewicz[1,6,*]

[1]Department of Materials Science and Chemical Engineering, Stony Brook University, Stony Brook, NY 11794
[2]Department of Nuclear Engineering, University of New Mexico, Albuquerque, NM 87131
[3]National Synchrotron Light Source II, Brookhaven National Laboratory, Upton, NY 11793
[4]Department of Nuclear Engineering, University of Tennessee, Knoxville, TN 37996
[5]Department of Materials Science and Engineering, Northwestern University, Evanston, IL 60208
[6]Institute for Advanced Computational Science, Stony Brook University, Stony Brook, NY 11794

*Corresponding Author: jason.trelewicz@stonybrook.edu

## Abstract

A large body of literature within the additive manufacturing (AM) community has focused on successfully creating stable tungsten (W) microstructures due to significant interest in its application for extreme environments. However, solidification cracking and additional embrittling features at grain boundaries have resulted in poorly performing microstructures, stymying the application of AM as a manufacturing technique for W. Several alloying strategies, such as ceramic particles and ductile elements, have emerged with the promise to eliminate solidification cracking while simultaneously enhancing stability against recrystallization. In this work, we provide new insights regarding the defects and microstructural features that result from the introduction of ZrC for grain refinement and NiFe as a ductile reinforcement phase – in addition to the resulting thermophysical and mechanical properties. ZrC is shown to promote microstructural stability with increased hardness due to the formation of $ZrO_2$ dispersoids. Conversely, NiFe forms into micron-scale FCC phase regions within a BCC W matrix, producing enhanced toughness relative to pure AM W. A combination of these effects is realized in the WNiFe+ZrC system and demonstrates that complex chemical environments coupled with the tuning of AM microstructures provides an effective pathway for enabling laser AM W materials with enhanced stability and performance.

## 1. Introduction

Refractory metals (tungsten, molybdenum, tantalum, rhenium, etc.) have attractive properties for applications in extreme operating environments containing high thermal and radiation fluxes such as those experienced by plasma-facing materials in existing and future fusion energy devices [1-3]. While tungsten (W) has emerged as the leading candidate in planned fusion demonstrations [4], it continues to suffer from a high ductile-brittle transition temperature (DBTT) and can be further embrittled by exposure to irradiation [5-9]. These issues undermine its effectiveness by reducing machinability and limiting operation temperature. Furthermore, W recrystallizes at intermediate elevated temperatures (i.e., 1100-1300 °C depending on material grade), leading to a reduction in the high temperature strength [10], and raising the DBTT [11]. Alloying of W has been shown to alter the impurity distribution in the W matrix, which is often correlated with ductility and recrystallization resistance. Several different approaches have led to enhanced performance metrics relative to pure W, and some of these improvements offer distinct benefits for mitigating crack formation in additively manufactured W-based materials.

Recent work on W-Re and W-K alloys have demonstrated marked improvements in low and high temperature performance, respectively. W-Re alloys, from 1-25 at.% Re [12], lower the DBTT, increase both toughness and ductility [13], and have limited recrystallization [14]. K-doped W, typically employed in high temperature applications, relies on the formation of nanoscale K bubbles to kinetically pin W grains during high temperature operation, thereby preventing grain growth to preserve the W microstructure [15-17] and limiting creep [18]. Some oxide and carbide dispersion-strengthened materials have also demonstrated enhanced thermal stability due to the same mechanisms, but interestingly with increased ductility at fusion-relevant temperatures [19-26]. With these approaches relying on the kinetic pinning of grain boundaries (rather than alternative thermodynamic approaches [27-29]), it is intrinsically limited at high homologous temperatures due to the Arrhenius temperature dependence of the grain boundary mobility [30].

Other approaches to alloying W instead utilize elements that phase separate from W and form a ductile matrix. This class of materials is often referred to as tungsten heavy alloys and generally contain additions of Ni, Fe, Co, and/or Cu in concentrations that maintain W as the solvent. At alloying concentrations of up to ~10 wt.%, improved sinterability and ductility [31] have been reported, albeit with a reduced density and recrystallization temperature [32, 33]. As with many of the aforementioned alloys, heavy alloy synthesis has largely focused on conventional processing methods (arc melting or sintering), which require long duration exposure to high temperatures and result in coarse-grained samples and inherent constraints on fabricating complex shapes without significant post-processing.

Additive manufacturing (AM) encapsulates a relatively new class of energy-beam fabrication techniques [34, 35], which allow for near-net shape material fabrication and direct process control pathways while eliminating post-fabrication material working. Due to limitations in hot forging geometries, AM techniques such as Laser Powder Bed Fusion (L-PBF), have been recently employed to produce W parts with complex geometries [36-40]. However, AM W is riddled with cracks due to the affinity of W to form embrittling features at grain boundaries upon solidification (e.g., oxides and pores), repeated heating and cooling through its DBTT during AM, and large thermal expansion strains [41, 42]. Crack formation is attributed to hot cracking, an umbrella term from welding to describe cracks that form at elevated temperatures [42, 43]. While typically associated with solidification, the driving force is defined as shrinkage strains between grains [44, 45] with tungsten being particularly susceptible due to its low fracture toughness and high DBTT (~723 K). On this basis, Vrancken et al. [42, 46] employed *in situ* high-speed video to demonstrate that part heating above the DBTT can eliminate cracking in tungsten plates and rods, but concluded that this strategy has limited effectiveness in AM processing due to enhanced impurity diffusion in high surface area powders. Other strategies from the welding community center on reducing these shrinkage strains by altering the melt pool shape, reducing impurities, and/or grain size refinement.

Owing to these issues, AM of W alloys have been investigated [37, 47], with several W-Ta [48] and W-ZrC [49] alloys exhibiting reduced cracking relative to commercially pure AM W. Iveković et al. [50] considered a commercial W heavy alloy (W-7Ni-3Fe, wt. %), hypothesizing that these elements will increase the temperature over which solidification occurs and enable the liquid to accommodate grain shrinkage. They also examined subsequent heat treatments to liquify the NiFe phase, which filled existing cracks and led to ductility in tension [50]. Unlike the AM W-Re alloys, where cracks in solidified components can only be healed after two high temperature heat treatments – one under a hydrogen atmosphere at 2400 °C and the other under a pressure of 200 MPa at 2000 °C [51] – these heat treatments can be carried out at much lower temperatures and for shorter times. Alloying therefore represents a promising pathway for the elimination of processing-induced cracking in AM tungsten. However, there are several outstanding questions regarding the interplay between elemental constituents, their concentration, and the resulting microstructure during solidification.

In this study, we examine microstructures in AM W alloys, fabricated via L-PBF, with the alloying additions chosen based on their potential to reduce/mitigate solidification crack formation. Specifically, ZrC is added to W to refine the microstructure during solidification, while a ductile FCC NiFe phase is explored as a toughening element to help accommodate large thermal strains. Combined effects of these two alloying additions are also considered in the multicomponent WNiFe+ZrC system. Microstructure and phase distribution in each alloy are characterized through electron microscopy and multimodal X-ray analysis methods. Our results provide definitive proof that carbide dispersions reduce crack density via grain refinement, while NiFe additions effectively backfill solidification cracks but at the expensive of mass loss via vaporization, which enhances the retained porosity. In addition, the FCC NiFe phase is shown to be integrated into the W grain structure unlike traditional heavy W alloys where it serves as a matrix between primary W particles. These resulting microstructures provide insights into the scaling of hardness and thermal properties with alloy composition and subsequent thermal aging.

## 2. Materials and Methods

Four alloys were considered in this study: W, W+0.5ZrC, W-3.5Ni-1.5Fe, and W-3.5Ni-1.5Fe+0.5ZrC. W powders (-325 mesh) were acquired from Global Tungsten & Powders, Ni and Fe powders (-325 mesh) were acquired from Fisher Scientific, and 20 nm ZrC nanopowders were acquired from US-Nano. Ni and Fe powders were added to W via mechanical tumbling for 3-8 hours, and ZrC was introduced via high-energy ball milling for 1-2 hours. The heavy alloy composition is more W-rich than the study by Iveković et al. [50] but should still take advantage of the large solidification range afforded by Ni+Fe. The small fraction of ZrC was chosen to both act as an oxygen getter and a grain refiner during solidification.

Cylindrical specimens, 12.5mm in diameter and 10 mm in height, were printed on steel build plates using a commercial EOS L-PBF machine. In this study, the goal was not to optimize laser processing parameters but rather have samples built under similar conditions for comparison. A range of volumetric energy densities was considered following the work in Iveković] et al. [50], where energy densities between 175-200 Jmm$^{-3}$ provided sufficiently dense specimens across all four compositions. The densities determined via the Archimedes method were as follows: W (15.7 gcm$^{-3}$), W+0.5ZrC (14.8 gcm$^{-3}$), W-3.5Ni-1.5Fe (16.5 gcm$^{-3}$), and W-3.5Ni-1.5Fe +0.5ZrC (16.2 gcm$^{-3}$). Cylinders were cut via electrical discharge machined, 2.0 mm above the build plate, to avoid chemical mixing. Samples were later sectioned into thin disks ~1 mm in height for thermal property characterization and hardness testing. Using a Netzsch 467HT Laser Flash Analyzer (LFA) apparatus, thermal diffusivity was determined from room temperature to 1000 °C in 200 °C increments on bulk cylindrical samples with diameters of 12.5 mm and thicknesses of 2 mm. A graphite standard reference was used to determine the thermal conductivity. Vickers indentation was performed using a 500 gf load with reported hardness values determined from a minimum of 10 indents per sample.

A JEOL IT300 scanning electron microscope (SEM) operating at 30 kV was used for SEM energy-dispersive X-ray spectroscopy (SEM-EDS) elemental mapping to establish bulk composition distribution and locations for transmission electron microscopy (TEM) sample fabrication. An EDAX Octane Elite Super EDS detector and EDAX Velocity Super electron backscatter diffraction (EBSD) camera coupled to the IT300 SEM was used for SEM-EDS and SEM-EBSD mapping, respectively. Samples for TEM were prepared through a typical focused ion beam (FIB) milling and lift-out procedure using a Scios 2 DualBeam SEM-FIB operated with a Ga beam at 30 keV with final cleaning steps at 5 and 2 keV. The I$^3$TEM at Sandia National Laboratories, a JEOL 2100 TEM operated at 200 kV, was used for bright-field TEM (BF-TEM) imaging. Automated Crystallographic Orientation Mapping (ACOM) was performed on the JEOL 2100 using the Nanomegas ASTAR analysis software. An 11 nm step size and 10 cm camera length were utilized to collect the diffraction patterns as the beam was rastered across the TEM samples. The Topspin Software (Nanomegas SPRL) package was used to collect the diffraction patterns with matching of the collected diffraction patterns to the correct phase accomplished using the Index2 software (Nanomegas SPRL) package. Further clean-up of the ACOM data was processed via TEAM software (Ametek, Inc.), performing a nearest neighbor confidence index and orientation cleanup recipes that changed <5% of the indexed data points. Further microstructural characterization via STEM-EDS was carried out with an FEI Talos F200X located in the Center for Functional Nanomaterials (CFN) at Brookhaven National Laboratory (BNL).

Two-dimensional scanning X-ray spectroscopy measurements were performed at the Hard X-Ray Nanoprobe (HXN) beamline at the National Synchrotron Light Source-II (NSLS-II) to extract elemental [52] and chemical [53, 54] information on the microstructural features observed in these systems. The incident X-rays had a wavelength of 1.3776 Å (9 keV) and were focused via a Fresnel zoneplate nanofocusing optics to a ~40 × 40 nm spot size employing scanning dwell times of 20-100 ms. The samples examined in this study were TEM lamellas of nominally 100 nm thickness. Nanoscale X-ray absorption near-edge spectroscopy (nano-XANES) [53] around the Fe and Ni *K-edges* (7112 and 8333 eV, respectively) were performed to resolve the local atomic structure within the different phase regions determined through STEM-EDS with a spatial resolution of 40 nm. Two-dimensional X-ray florescence (XRF) spectral maps (60 and 72 individual maps for Fe and Ni K-edges, respectively) were processed and visualized in PyXRF [55] and MIDAS [56]. Full-field X-ray computed tomography (XCT) data sets were collected for a series of specimens, employing a Zeiss Xradia 520 Versa X-ray microscope. The X-ray source was operated at 140kV and 71.6 µA (10W). The isotropic voxel size was set to 1.1 µm and a series of 800 projections were collected over 360° with a 10 s image collection time per step. Data reconstruction was performed using a filtered-back projection algorithm and visualized in ImageJ.

Bulk samples for the *ex situ* XRD annealing study were subjected to heat treatments at 1150 and 1350 °C for 5 hrs under a reducing gas atmosphere (Varigon H5, 95% Ar and 5% H) in an MTI GSL-1600X tube furnace. XRD measurements were performed at the X-ray Powder Diffraction (XPD) beamline at NSLS-II. All measurements were performed in transmission mode with an amorphous silicon-based flat panel detector (Perkin-Elmer) mounted orthogonal to, and centered on, the beam path. The sample-to-detector distance and tilts of the detector relative to the beam were refined using a $CeO_2$ (SRM 674b) powder standard. The wavelength of the incident X-rays was 0.1839 Å at 67.40 keV. The sample-to-detector distance was calculated to be 133.56 cm. Multiple patterns were acquired to ensure collection of isotropic diffraction rings with all raw two-dimensional (2D) XRD patterns background corrected by subtracting the dark current image and Kapton/air scattering. Noticeable artifacts in the 2D images (beam stop or dead pixels) were masked. The corrected and masked two-dimensional detector images were then averaged and radially integrated to obtain the one-dimensional XRD patterns.

The background subtracted XRD patterns were analyzed with the TOPAS software package (BRUKER) using a modified pseudo-Voigt function for fitting of the peak profiles. The instrument contribution to the broadening of the measured profiles was quantified by fitting a $CeO_2$

NIST powder standard, with known crystalline-domain size and negligible strain contribution. The Gaussian and Lorentzian-based broadening parameters were subsequently fixed during the analysis of the alloys under investigation. The weight fraction, coherent domain size, lattice parameter, and microstrain for the body-centered cubic (BCC) W matrix phase were allowed to vary during the refinements. The microstrain components for the minor face-centered cubic (FCC) NiFeW phase (present in WNiFe and WNiFe+ZrC) were not included in the refinements due to the known complications associated with their quantification at such small sizes (the refined size parameters of the FCC phase are thus lower limits).

## 3. Results

### *3.1. Microstructure and Morphology*

X-ray computed tomography was performed to determine the extent of porosity and cracking formed during processing; representative slices through the three-dimensional volumes of each microstructure are shown in **Figure 1**. Extensive cracking is observed in pure W, as expected for AM W microstructures [41]. The addition of ZrC reduced crack formation in the microstructure; however, a higher number of pores are observed compared to pure W. Extensive porosity is observed in both W heavy alloy microstructures, with a higher degree of porosity in WNiFe relative to WNiFe+ZrC. The XCT results in **Figure 1** show that alloying can reduce cracking at the expense of increasing porosity, but the latter can be more easily addressed through process optimization and/or heat treatment [57]. However, the combination of the NiFe alloying additions with ZrC particles collectively mitigated cracking while reducing the degree of porosity.

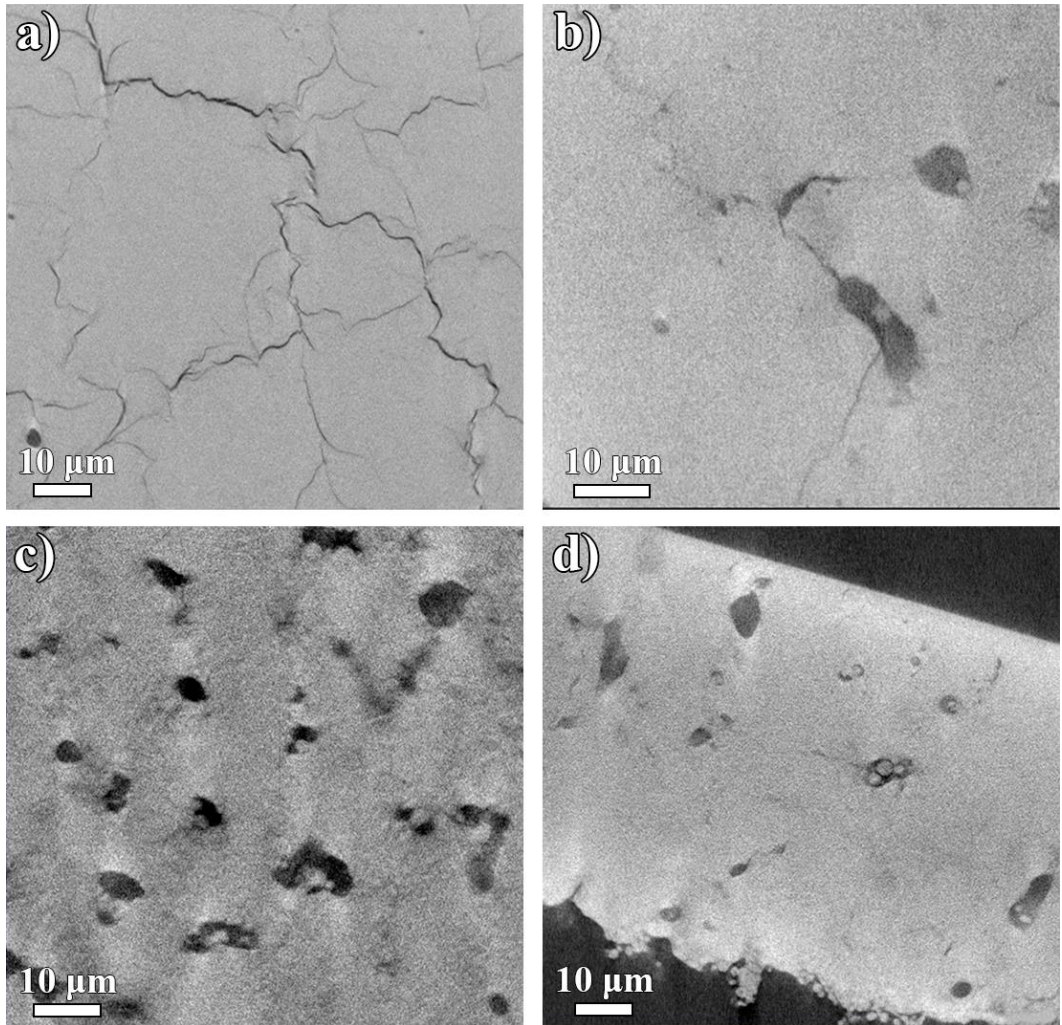

*Figure 1.* X-ray computed tomography slices through 3D volumes of as-printed AM (a) W, (b) W+ZrC, (c) WNiFe, and (d) WNiFe+ZrC. Cracking in the AM W and W+ZrC is eliminated in the WNiFe and WNiFe+ZrC samples but at the expense of porosity.

Low magnification SEM micrographs of the sample microstructures and corresponding elemental composition maps are shown in **Figure 2** for the commercially pure W and alloys of W+ZrC, WNiFe, and WNiFe+ZrC. Macroscale cracking (indicated by the black arrows) and pores (indicated by the white arrows) are evident on the surface of the W and W+ZrC samples. Porosity is evident in the ternary heavy W alloy (both WNiFe and WNiFe+ZrC), in addition to spherical W-rich regions (indicative of unmelted W powder particles) and a secondary phase at the W interfaces. SEM-EDS maps of the secondary phase reveal that they are predominantly composed of Ni and Fe. Furthermore, the distribution of Zr varies significantly between W+ZrC and WNiFe+ZrC; in the W+ZrC sample, Zr appears to segregate into 1-10 µm particles at W grain boundaries while in the WNiFe+ZrC sample, it is homogeneously distributed.

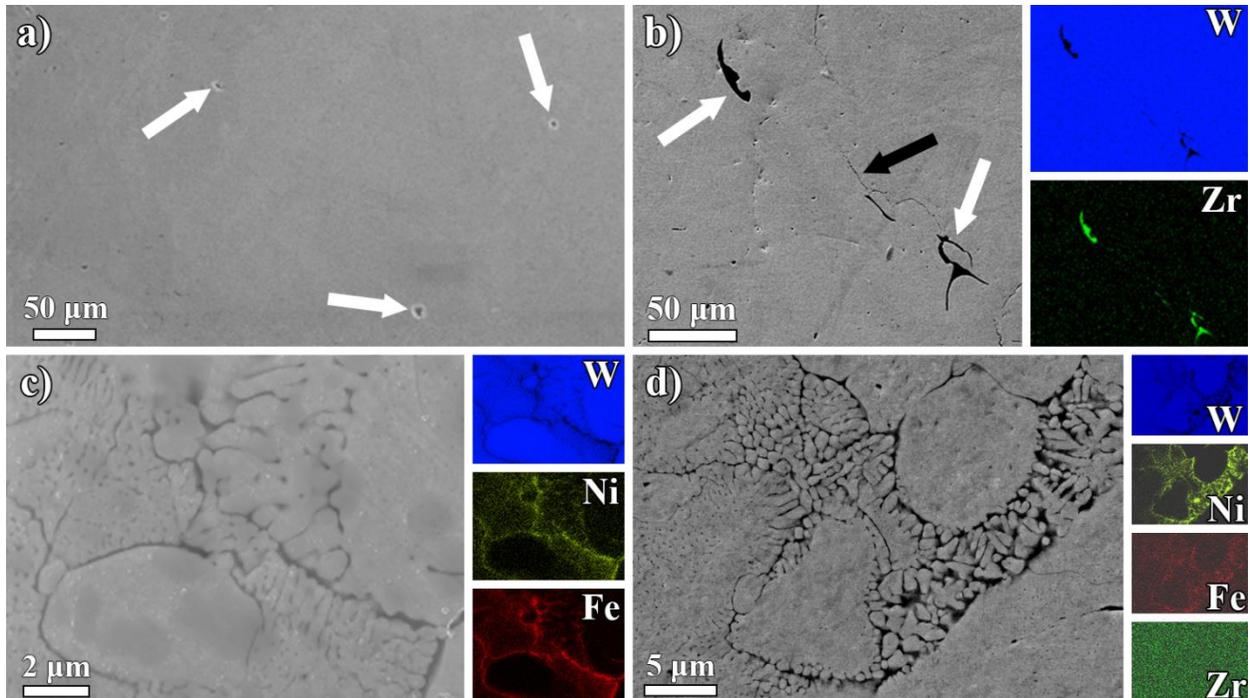

*Figure 2. Secondary electron SEM micrographs and corresponding SEM-EDS elemental maps for samples: (a) W (No EDS mapping given the absence of alloying elements), (b) W+ZrC, (c) WNiFe, and (d) WNiFe+ZrC. Porosity is evident in the samples (indicated with white arrows), while cracking is evident in the W+ZrC sample (indicated with black arrows). The WNiFe and WNiFe+ZrC samples have NiFe-rich regions, appearing in black, distributed intergranularly between W grains, including some spherical unmelted W grains.*

The crystallographic orientations for each microstructure are shown in **Figure 3**, through out-of-plane EBSD inverse pole figure (IPF-Z) maps of grains indexed as BCC W. Pure W exhibits a typical AM solidification microstructure, where grain topologies follow melt-pool boundaries [41]. However, the microstructure changes drastically with the addition of the secondary phase elements. ZrC (**Figure 3**b) refines the grain structure from tens of micrometers for pure W to ultra-fine grain sizes in W+ZrC. At these temperatures, ZrC will melt and either restricts the grain size during solidification or via Zener pinning – or a combination of both [58-61]. The addition of NiFe produces a heterogeneous grain microstructure with more equiaxed grains that are varied in size. Some spherical W grains are observed (similarly to the SEM micrographs in **Figure 2**), indicative of unmelted W powder particles. Finally, the WNiFe+ZrC sample demonstrates substantial texturing, on a scale of tens of micrometers, with ultra-fine grains.

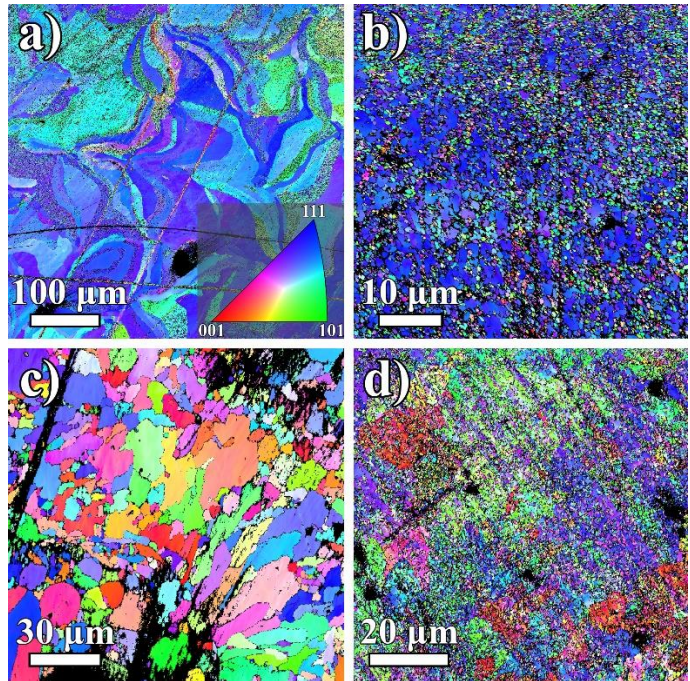

*Figure 3. EBSD IPF-Z maps of the grain structure of as-printed AM (a) W, (b) W+ZrC, (c) WNiFe, and (d) WNiFe+ZrC.*

STEM high-angle annular dark-field (STEM-HAADF) micrographs of the local microstructure for the W and W+ZrC samples are provided in **Figure 4**. The W lamella (**Figure 4**a) consisted primarily of a single grain, with a grain boundary evident at the bottom right of the lamella. The W+ZrC sample under STEM-HAADF imaging conditions (**Figure 4**b) consisted of the W matrix (region with lighter contrast) and a Zr dispersoid (region with darker contrast), consistent with the SEM-EDS map in **Figure 2**b. The Zr dispersoid is approximately 1 µm wide and has a sharp interface with the W matrix; this dispersoid likely contains twins, evidenced by the numerous parallel lines and corresponding contrast features in **Figure 4**b. STEM-EDS maps reveal W-rich and Zr-rich regions, with the Zr-rich region containing appreciable O-content. There is no evidence of ZrC, which is not surprising because cubic $ZrO_2$ twins easily and ZrC may be decomposed into Zr and C during the AM fabrication process, at which point the Zr readily oxidizes into $ZrO_2$.

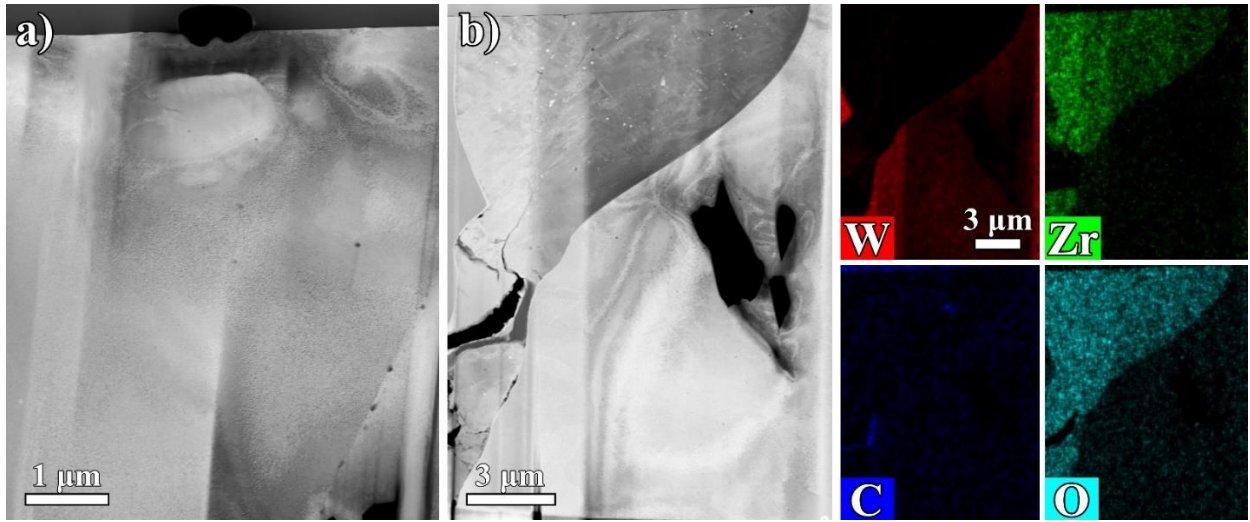

*Figure 4. STEM-HAADF micrographs of as-printed (a) W and (b) W+ZrC with STEM-EDS maps of W, Zr, C, and O acquired on a region containing a grain boundary in the W+ZrC sample.*

STEM-HAADF images for the WNiFe sample are shown in in **Figure 5**a. The large variations in the enhanced Z-contrast suggestive of a two-phase microstructure [62] with a dendritic component as observed in **Figure 2**c. The lighter phase is consistent with a predominantly W matrix while the second phase is significantly darker in contrast and rich in the lighter elements (i.e., Ni and Fe). These regions are found primarily between W grains and are < 500 nm in size with sharp interfaces and free of porosity. STEM-EDS maps indicate a correspondence between Ni and Fe in these second phase regions, denoted as NiFeW for the remainder of this text, with a total concentration of 45 at.% W, 39 at.% Ni, and 16 at.% Fe (compared to the BCC W region containing 98 at.% W, 1 at.% Ni, and 1 at.% Fe).

The IPF-Z ACOM map is also shown, indexed for BCC W (red) and FCC Ni (green) grains, and confirms the presence of a two-phase microstructure as suggested by the STEM-HAADF/EDS analysis. The WNiFe+ZrC sample also exhibited NiFeW-rich regions distributed within a W-rich matrix as evidenced by the Z-contrast in the STEM-HAADF image in **Figure 5**b and confirmed in the EDS maps. These NiFeW regions are akin to those in the ternary alloy and nominally 200 nm wide. However, a significantly larger number of interfaces was found in the WNiFe+ZrC sample, which is attributed to the addition of ZrC promoting the formation of a finer microstructure during solidification (**Figure 3**). The STEM-EDS maps also reveal the affinity for Zr to occupy the FCC NiFeW phase, but with a subtle gradient into the BCC W phase especially around the phase boundaries. Finally, the ACOM map in the inset confirmed the two-phase microstructure, which relative to the WNiFe sample free of Zr, exhibited a more defined FCC NiFeW phase with structurally sharp interfaces despite the grading of Zr across the phase boundaries.

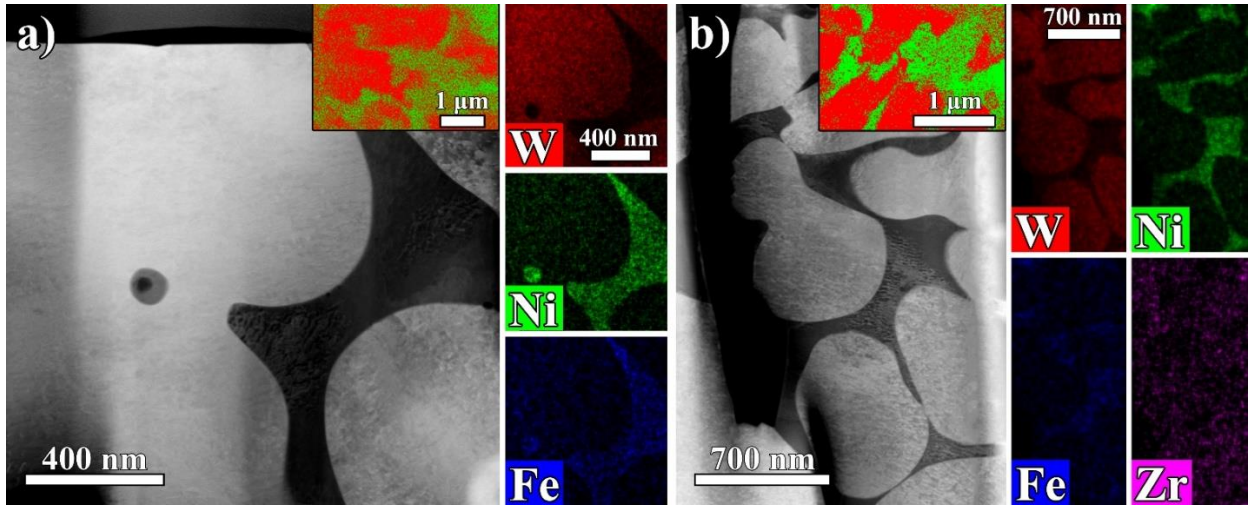

*Figure 5. STEM-HAADF micrographs and STEM-EDS maps (W, Ni, Fe, and Zr) of as-printed (a) WNiFe and (b) WNiFe+ZrC. ACOM phase maps are provided as insets.*

### 3.2. Phase and Electronic Structure

For the samples containing NiFe additions, nano-XANES measurements were conducted around the Fe and Ni K-edges (7112 and 8333 eV, respectively) to "fingerprint" the local structure within the different phase regions. Results for the WNiFe specimen are shown in **Figure 6** with the 2D XRF maps at both K-edges shown in the insets revealing the two-phase microstructure including the bulk W region (mapped in blue) containing low concentrations of Ni and Fe and secondary region with high concentrations of Ni and Fe (mapped in green/yellow), consistent with the STEM-EDS maps in **Figure 5**a. Examination of the XANES spectra collected at the Ni K-edge (**Figure 6**a) shows evidence of two features: (i) the NiFeW phase (black spectrum collected from the region indicated with a black square) is predominately FCC, as no notable difference is present when compared with the FCC Ni standard (blue spectrum), and (ii) the bulk region (red spectrum from the region indicated with a red square) corresponds to a BCC W phase with a trace concentration of interstitial Ni atoms. Relative to the FCC Ni standard, the XANES spectra of the Ni atoms within the bulk W-phase exhibited a reduced pre-edge feature (~8340 eV), a more pronounced post-edge peak (8351eV), and additional multiple scattering peaks at 8365 and 8370 eV. At the Fe K-edge in **Figure 6**b, the XANES spectra attributed to the FCC NiFeW phase (black) is consistent with the spectra observed at the Ni K-edge. The XANES spectra for the Fe atoms within the bulk BCC W phase (red) matches the BCC Fe standard (blue) closely at the pre-edge feature and exhibits a similar post-edge peak and an additional peak at 7160 eV.

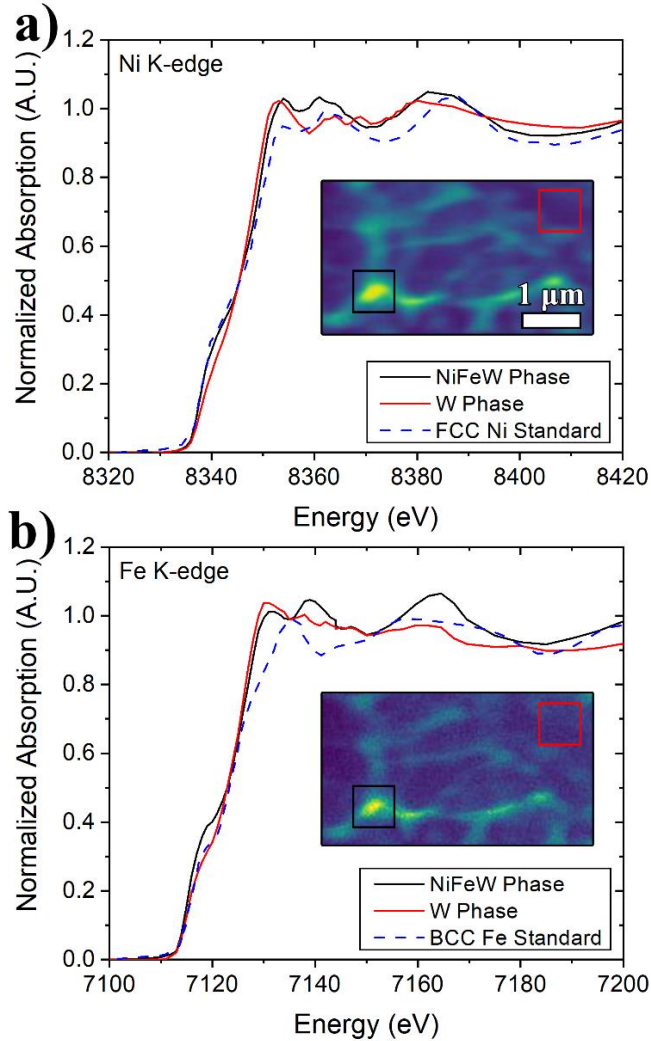

*Figure 6.* XANES spectra from selected regions in AM WNiFe at the (a) Ni K-edge (8333 eV) and (b) Fe K-edge (7112 eV). Composite 2D XRF maps are provided as insets with the corresponding XANES spectra from the identified regions matching the respective colors denoting the measurement locations. FCC Ni and BCC Fe XANES standards are provided for reference [63].

Differences in the shape and intensity of features around the K-edge are linked to the electronic structure and atomic environment of the absorbing and neighboring atoms [64, 65]. For example, prominent features in the Ni and Fe XANES regions result directly from multiple scattering resonances of the $1s$ photoelectron in the continuum [66, 67]. The differences in the Fe and Ni K-edge spectra quantified from the W matrix and NiFeW regions demonstrate clearly the different atomic environments. Specifically, substitutional Ni and Fe atoms are present within the W BCC matrix while the NiFeW segregated regions exhibit an FCC crystal structure aligned with the Ni standard. Similar results are observed in the XANES spectra collected at both Ni and Fe K-edges for the WNiFe+ZrC sample in **Figure 7**. The inset XRF maps reveal two distinct phases with the BCC W phase exhibiting the reduced pre-edge feature and identical features in the post-edge region (shown in red). The second phase corresponds to FCC NiFeW, with a noticeably large pre-edge feature and four peaks in the post-edge region (shown in black) at 8352, 8359, 8362, and 8381 eV. The 2D XRF maps and XANES results indicate that the addition of the ZrC does not significantly influence the microstructural phase of the other two components despite the difference in Zr distribution from in the STEM-EDS maps.

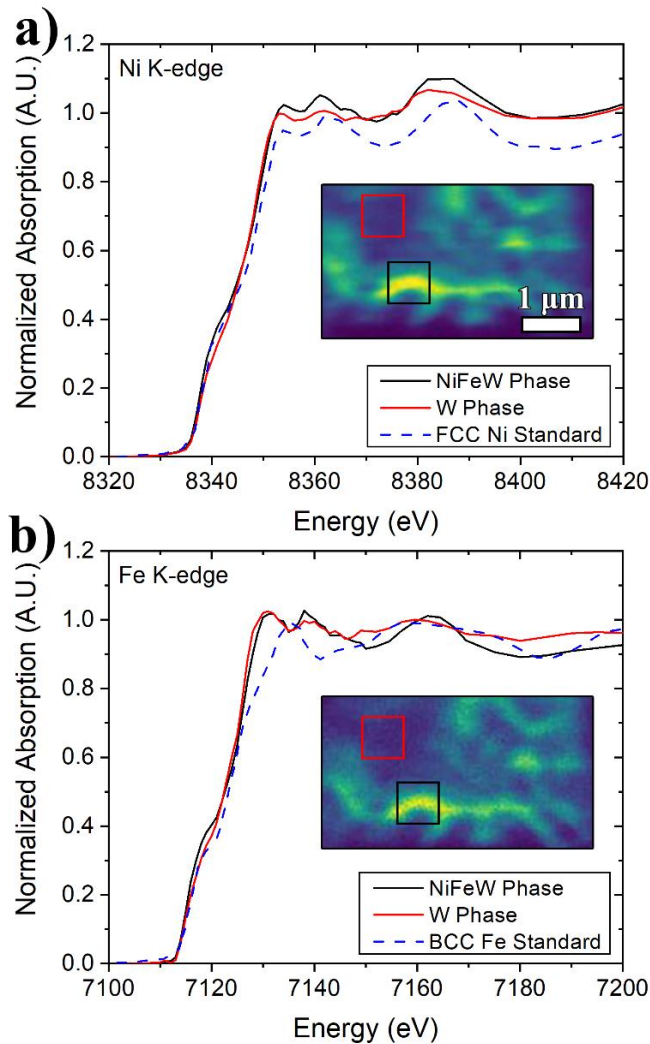

*Figure 7. XANES spectra from selected regions in AM WNiFe+ZrC at the (a) Ni K-edge (8333 eV) and (b) Fe K-edge (7112 eV). Composite 2D XRF maps are provided as insets with the corresponding XANES spectra from the identified regions matching the respective colors denoting the measurement locations. FCC Ni and BCC Fe XANES standards are provided for reference [63].*

Synchrotron XRD patterns for the as-printed W and W alloy samples are shown in **Figure 8** with crystallographic indices corresponding to BCC W and FCC Ni indexed where appropriate. The unalloyed W exhibited only BCC W, as expected, which was also the dominant phase in the W+ZrC sample. Given the twinning apparent in the Zr-rich phase of the W+ZrC microstructure (as shown in **Figure 4**b), it is likely that the corresponding Zr phase peak is too broad and weak to be distinguished from the background. The XRD patterns acquired on the WNiFe and WNiFe+ZrC samples exhibited peaks corresponding to a secondary FCC Ni phase in addition to the primary BCC W phase. From the nano-XANES maps in **Figure 6** and **Figure 7**, these peaks derive from the NiFeW inclusions without the appearance of any additional Zr-based phases, consistent with Zr being dissolved in the FCC NiFeW solid solution from **Figure 5**b.

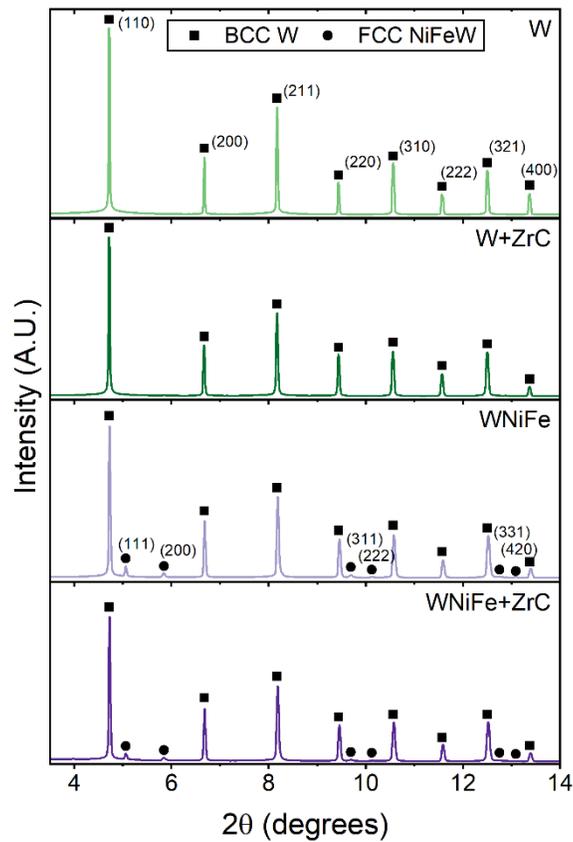

*Figure 8. XRD data of all as-printed bulk AM samples with BCC W phases indexed with a square and FCC phases with a circle.*

Lattice parameter, coherent domain size, microstrain, and phase fractions quantified from the XRD patterns are provided in **Table 1**. Lattice parameters across all the printed samples are reduced relative to typical values for BCC W, e.g., 3.165 Å [68] with smaller contractions (~.005

Å) in W and W+ZrC and larger contractions (~.01 Å) in WNiFe and WNiFe+ZrC, a likely reflection of alloying in the BCC W phase The larger lattice parameter for the FCC NiFeW phase is in agreement with the STEM-EDS maps suggesting that small amounts of W are dissolved into the FCC lattice and consistent with previous observations in other W heavy alloys [69-71]. The increased compositional complexity also manifested in the microstructural features accessible from XRD, specifically the microstrain and coherent domain size also captured in **Table 1**.

*Table 1. Lattice parameter, X-ray coherent scattering domain size, microstrain, and phase fractions as determined via XRD for the as-printed bulk AM W samples.*

| Sample Composition | Phase | Lattice Parameter (Å) | Coherent Domain Size (nm) | Microstrain (A.U.) | Phase Fraction (wt. %) |
|---|---|---|---|---|---|
| W | BCC W | 3.16047 ± 0.00006 | 654.0 ± 76.0 | 0.010 ± 0.001 | - |
| W+0.5ZrC | BCC W | 3.16162 ± 0.00004 | 96.4 ± 3.5 | 0.023 ± 0.001 | - |
| W-3.5Ni-1.5Fe | BCC W | 3.15541 ± 0.00006 | 59.1 ± 1.2 | 0.029 ± 0.002 | 96.2 |
| W-3.5Ni-1.5Fe | FCC Ni | 3.60980 ± 0.00048 | 20.8 ± 1.0 | - | 3.7 |
| W-3.5Ni-1.5Fe +0.5ZrC | BCC W | 3.15604 ± 0.00005 | 57.4 ± 1.0 | 0.044 ± 0.002 | 97.2 |
| W-3.5Ni-1.5Fe +0.5ZrC | FCC Ni | 3.60875 ± 0.00088 | 16.5 ± 1.0 | - | 2.8 |

The microstrain was nominally low in pure W with a large coherent domain size, consistent with the general lack of cellular dislocation networks in laser AM W due to the low thermal expansion of W, as the primary source of dislocations in AM metals is generally attributed to plastic deformation induced by thermal expansion/shrinkage in a constrained medium [72]. The alloys all exhibited greater microstrain relative to pure W, with comparable increases upon the addition of ZrC and NiFe to 0.023 and 0.029, respectively. Combining these two alloying additions in the WNiFe+ZrC samples produced a further increase in the microstrain relative to the ternary alloys to 0.044, indicating that the microstructures became increasingly strained in the presence of more complex alloying environments due to the increased solute concentrations in the bulk BCC W phase and increased grain boundary densities [73]. Accompanying this change in microstrain with added solute concentrations (especially relative to the unalloyed W sample) were reductions in the coherent domain sizes, where the addition of ZrC dropped this length scale relative to pure W with a further decrease for the WNiFe alloy but with little additional change upon the addition of ZrC to this ternary. Collectively, the finer coherent scattering sizes and concomitant increases in microstrain indicate that the addition of solute elements produced more complex crystalline microstructures containing a higher degree of defects. We note that the coherent domain size determined from XRD does not reflect the actual grain sizes of the samples, as the presence of higher order defects has been shown to significantly reduce coherent scattering especially in AM materials [54, 74].

### *3.3. Thermal Stability and Thermophysical Properties*

Thermal stability was investigated at two temperatures of 1150 °C and 1350 °C, selected based on common recrystallization temperatures for tungsten [75] and analyzed via *ex situ* XRD to capture the evolution of atomic and microstructural parameters. XRD patterns (not shown) for the annealed samples appeared qualitatively identical to the patterns for the corresponding as-printed condition in **Figure 8** with Rietveld refinements employed to extract subtle changes in peak positions and their degree of broadening. Microstrain was quantified for the primary BCC W phase with relative phase fractions considering the FCC NiFeW phase, and results are shown in **Figure 9**. Relaxation of the microstructure manifested as a reduction in the microstrain during annealing in **Figure 9**a with W and WNiFe exhibiting a monotonic decrease as a function of annealing temperature with terminal average values of 0.005 and 0.020 at 1350 °C, respectively. The W+ZrC sample also exhibited microstructural relaxation, but with a large reduction upon annealing at 1150 °C and only a little additional change following the 1350 °C heat treatment. Conversely, the microstrain remained approximately constant, within error, for the WNiFe+ZrC sample. Accompanying the stability exhibited by this sample was a consistent phase distribution through annealing, where the fraction of the FCC NiFeW phase remained at approximately 3 wt. % from **Figure 9**b. The fraction of the FCC NiFeW phase in the WNiFe sample also remained constant at approximately 3.7% regardless of the annealing temperature.

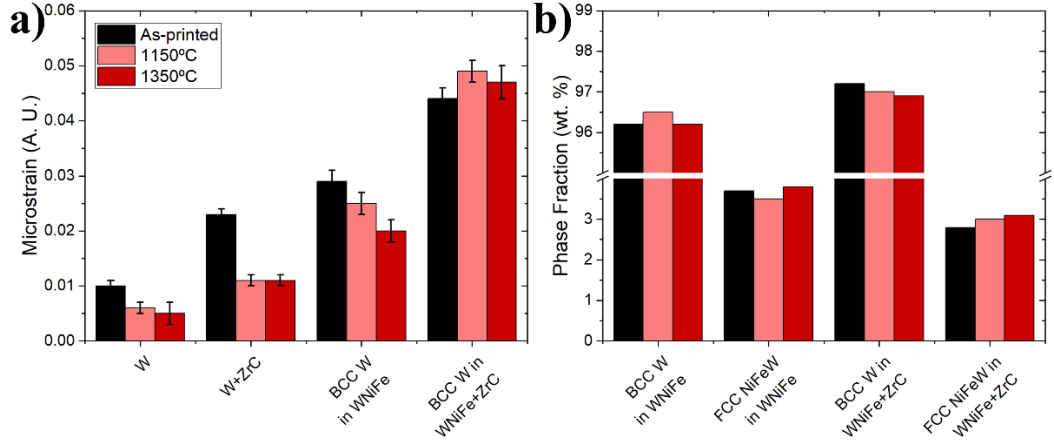

***Figure 9.*** *(a) Microstrain and (b) phase fraction for all AM W microstructures in the as-printed and annealing conditions.*

For the BCC W phase, in all specimens, values for the lattice parameter determined through a Rietveld refinement shown in **Figure 10**a were contracted relative to typical values for W (e.g., 3.165 Å). Annealing of the microstructures resulted in minor expansions of the lattice parameters: in W and W+ZrC the lattice parameter increased by nominally 0.04%, while in WNiFe and WNiFe+ZrC the lattice parameter increased by approximately 0.2%. Furthermore, the increase in the lattice parameter of BCC W in the heavy ternary alloys coincided with a substantial contraction in the lattice parameter of FCC NiFeW, as shown in **Figure 10**b, of nominally 0.5%. Thus, while relaxation of the microstructures is evident across all samples, this behavior is exacerbated in the heavy ternary alloys due to contraction of the secondary FCC NiFeW phase, which is attributed to the diffusion of Fe from the FCC NiFeW phase into the BCC W phase.

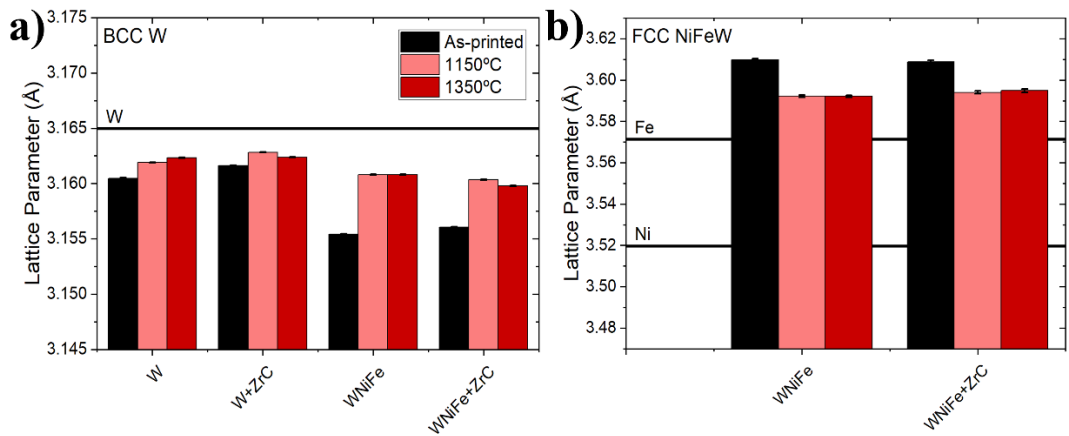

***Figure 10.*** *Lattice parameter for all (a) BCC and (b) FCC AM W phases in the as-printed and annealing conditions. Literature values for the lattice parameter of BCC W, FCC Ni, and FCC Fe from [76].*

The following behavior can be inferred from the XRD trends through annealing. First, in pure W, the absence of solute-rich heterogeneities eliminates such barriers to defect motion, and the reduction in microstrain is therefore a consequence of increased defect recovery at elevated temperature. Second, solute redistribution is not expected to occur in W+ZrC with temperature as Zr exists as $ZrO_2$. Instead, the refined grain sizes in W+ZrC allow for the rapid annihilation of 2D defects (e.g., dislocations, stacking faults, twins) at elevated temperature with no subsequent change in microstructure due to stabilization forces from the oxide dispersoids following the Zener pinning effect [77]. Third, solute distributions in the W heavy alloys contribute to higher lattice strains in the BCC W phase that are captured in the increased microstrain. In WNiFe, microstructural relaxation occurs during annealing following solute redistribution and concomitant enhanced mobility of the as-printed defects, which manifests as a gradual and continuous decrease in the microstrain with increasing temperature. Finally, in the case of WNiFe+ZrC, the addition of Zr in the bulk suppresses defect mobility relative to the ternary alloys, thereby stabilizing as-printed defects across the annealing temperatures considered in this study. It follows that microstructural stability in the printed alloys is a consequence of the increased number of obstacles to dislocation motion, which is directly connected to the increased microstructural complexity generated through alloying.

Given the minimal microstructural changes occurring above 1150°C, thermophysical properties were mapped at temperatures of up to 1000 °C to extract the thermal conductivity of the alloys relative to the printed pure W sample. Thermal conductivity, $\lambda$, was determined for each sample from the specific heat, $C_p$, and thermal diffusivity, $\alpha$, using the expression:

$$\lambda = \rho C_p \alpha, \tag{1}$$

where $\rho$ is the measured density. Thermal conductivity as a function of temperature is provided in **Figure 11**a for the W and W+ZrC samples relative to reference values for W [78]. Immediately evident are the significantly reduced thermal conductivities exhibited by the printed W and W+ZrC relative to the literature values for pure W. In the printed W, the disparity is due to the difference in microstructure and reduced density typical of AM materials. Furthermore, the addition of ZrC only served to further reduce thermal conductivity due to the presence of $ZrO_2$ dispersoids.

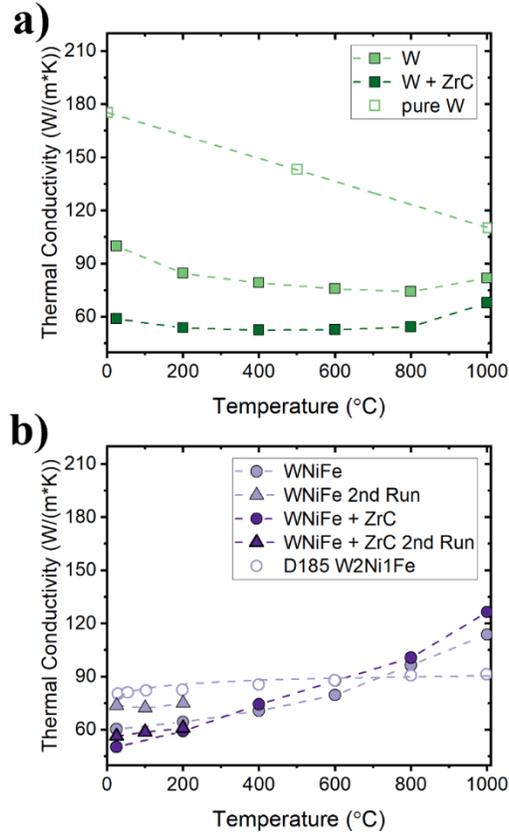

*Figure 11. Thermal conductivity as a function of temperature for (a) W and W+ZrC, and (b) WniFe and WniFe+ZrC samples. Given the solute redistribution determined from XRD analysis, the samples in (b) were heated twice, with thermal conductivity measured up to 200 °C for the second run. Thermal conductivity of pure W [78] and D185 W-2Ni-1Fe [79] are provided for reference.*

Thermal conductivity, and its scaling with temperature, for the WNiFe and WNiFe+ZrC alloys are shown in **Figure 11**b relative to values for a reference ternary D185 W-2Ni-1Fe alloy [79]. The trends were identical within measurement error for both samples with lower values relative to the reference WNiFe alloy below 600 °C and higher values above 600 °C. These alloys also aligned closely with the W+ZrC sample at temperatures up to 200 °C but diverged to larger values above this temperature. The absence of a plateau value for the AM WNiFe samples indicates that the microstructural changes through annealing were reflected in the scaling of the thermophysical properties with temperature. Samples from both WNiFe and WNiFe+ZrC were reheated to 200 °C following the initial heating cycle used to produce the trends in **Figure 11**. Trends up to this temperature for the second heating cycle provided in **Figure 11**b demonstrate that annealed WNiFe exhibited increased thermal conductivities relative to the as-printed values. Conversely, the thermal conductivity for WNiFe+ZrC did not exhibit any noticeable changes, signaling that the addition of ZrC promoted microstructural stability and is in agreement with the consistent microstrain values through annealing of this sample.

### *3.4. Insights into Alloying Effects on Mechanical Behavior*

Given the range of alloy chemistries and microstructural changes through annealing, the dependence of the mechanical properties on these factors was evaluated using indentation experiments. The indentation technique was deliberately selected to map the changes in hardness collectively with the implied toughness of the samples, where for brittle materials, microcracks generally form in lieu of residual pile-up from localized plasticity [80]. Vickers indents were performed on each sample in the as-printed condition and following annealing to 1150 and 1350 °C with results shown in **Figure 12**.

The average Vickers hardness for the as-printed W sample was 420 HV, consistent with prior hardness measurements on AM W [57, 81], with the addition of ZrC increasing it to 460 HV due to the presence of the $ZrO_2$ precipitates. The hardness of the as-printed WNiFe alloy was higher at 530 HV, despite the formation of the FCC NiFeW second phase, and is attributed to solid solution strengthening in both the BCC and FCC phases. Finally, the addition of ZrC to the WNiFe ternary further enhanced the hardness to 595 HV, which is due to solid solution strengthening of Zr homogeneously distributed in both FCC and BCC phases consistent with the STEM-EDS maps in **Figure 5**b. Representative surface morphologies around the residual impressions were also imaged for the as-printed samples and are shown in **Figure 12**(b-e). Both the W and W+ZrC samples exhibited clear evidence of microcracking, but with a reduced number of cracks generally observed around the residual impressions for W+ZrC. Nevertheless, any degree of cracking indicates that both the W and W+ZrC samples are brittle in the as-printed state, which agrees with reported indentation results on W [82]. Conversely, the periphery of the residual impressions produced on the WNiFe and WNiFe+ZrC samples were free of microcracking with evidence of pile-up on the surface of the WNiFe sample that appeared to align with the underlying dendritic microstructure in **Figure 12**d. These findings suggest an enhanced toughness in the as-printed WNiFe and WNiFe+ZrC samples relative to W and W+ZrC.

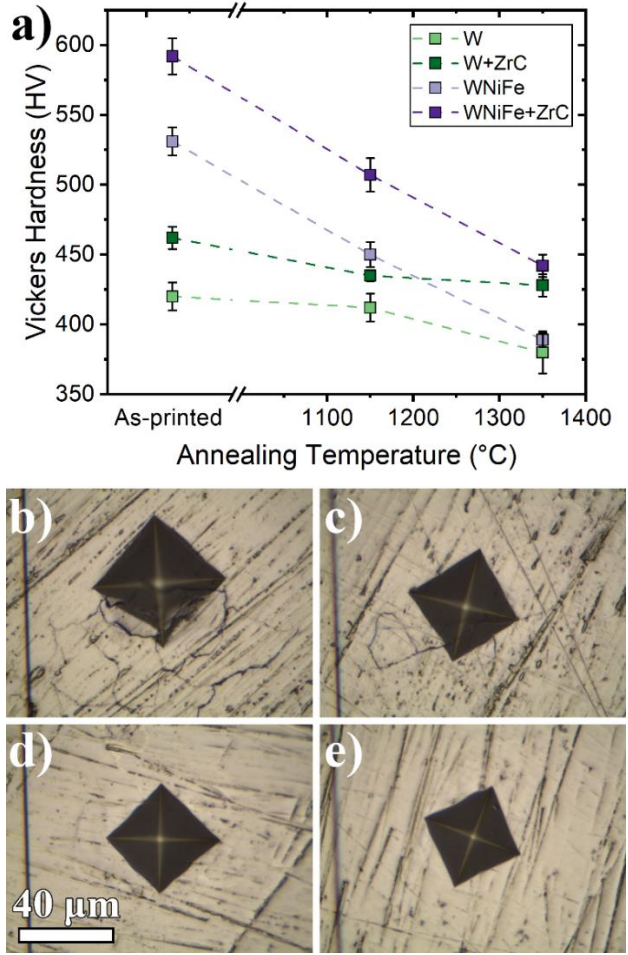

*Figure 12. (a) Vickers hardness for all AM W microstructures in the as-printed condition and annealed at the indicated temperatures. Optical micrographs of representative impressions on the as-printed samples of (b) W, (c) W+ZrC, (d) WNiFe, and (e) WNiFe+ZrC. Shared scalebar for all images provided in (d).*

Annealing led to reductions in the hardness across all samples but with differing temperature dependencies and overall degrees of softening. In W and W+ZrC, the decrease in hardness was minimal, decreasing to 415 and 440 HV after annealing at 1150 °C, respectively. However, the increase in annealing temperature to 1350 °C produced a marked decrease in the hardness for W to 390 HV while W+ZrC exhibited only a minor decrease relative to the 1150 °C condition. Given the temperature and time, the onset of recrystallization is expected in W and consistent with this decrease in hardness [83], thereby suggesting that the addition of ZrC – and subsequent decomposition to form $ZrO_2$ following laser AM – stabilized the alloy against recrystallization.

In the WNiFe microstructures, annealing produced significant and continuous reductions in the hardness, with the WNiFe alloy converging to the value for W while the WNiFe+ZrC sample approached the hardness of the W+ZrC sample. Unlike the W+ZrC sample, the decomposition of ZrC in the WNiFe ternary promoted Zr incorporation in both the BCC and FCC solid solutions rather than the formation of $ZrO_2$, thereby eliminating the oxide-induced stabilization mechanism exhibited by the W+ZrC alloy. Furthermore, the initial drop in hardness following the 1150 °C anneal aligns with the changes in the lattice parameter in **Figure 10**b, signaling that the changes in the degree of solid solution strengthening in each phase was responsible for softening over this temperature range. However, at 1350 °C, the lattice parameter plateaued, demonstrating that the additional reduction in hardness was not due to further solute redistribution but likely the onset of recrystallization in these alloys. Therefore, while the more complex alloys exhibited enhanced hardness and toughness due to the combination of solid solution strengthening and the presence of a more ductile FCC phase, temperature-induced microstructural instabilities lead to a precipitous reduction in strength relative to the as-printed condition with the upper temperature limit of the alloys inevitably constrained by the low melting point FCC NiFeW phase.

## 4. Discussion

Three unique additively manufactured W alloy microstructures were fabricated and characterized in this study with comparisons drawn to a pure AM W sample containing large (≥1 µm) irregular grains with melt-pool tracks and extensive microcracking in the as-printed state. The AM W prepared in this work exhibited a Vickers hardness (420 HV), which is comparable to typical polycrystalline W (400-450 HV [84]) and greater than recrystallized W (300 HV) [33], but with a high density of microcracks found in the periphery of the residual impressions as previously observed during indentation experiments on severely-deformed W [82]. Annealing at temperatures known to recrystallize deformed grades of tungsten [83] led to reductions in the hardness to below 400 HV (an approximately 10% reduction in hardness), indicating the W microstructure recrystallized above 1150 °C. The AM W sample also exhibited non-negligible reductions in thermal conductivity, with a value of 100 W/m·K at room temperature compared with ~164 W/m·K for polycrystalline W [78]. This behavior reflects the high crack density in the AM sample [85], shown in the XCT image of the microstructure shown in **Figure 1**a.

The first of the alloys – W+ZrC – exhibited fine grain sizes (≤1 μm) with the ZrC decomposing to form a distribution of $ZrO_2$ dispersoids with a smaller (i.e., down to 50 nm) population distributed within the W matrix and a larger (≥1 μm) population concentrated at interfaces. Interestingly, build tracks and the larger grain sizes reported by Li et al. [49] in AM W+ZrC were not immediately evident in this microstructure. The finer grain structure and coarse $ZrO_2$ dispersoids at interfaces more closely resembled spark plasma-sintered (SPS) W+ZrC composites [21], but with an additional population of small submicrometer oxides distributed within the W matrix – likely due to the fast solidification rate. The W+ZrC sample in this study combines attributes common to both AM and SPS W+ZrC microstructures, exhibiting a bimodal size distribution of $ZrO_2$ dispersoids with a finer grain microstructure. Solidification cracking was retained in this alloy despite the effective reduction in grain size and consistent with the distribution of $ZrO_2$ throughout the microstructure. Comparatively, the second alloy – WNiFe – also exhibited fined grain sizes (≤1 μm) but with an equiaxed microstructure containing two phase regions of BCC W and FCC NiFeW. This microstructure diverged from conventionally-sintered WNiFe heavy alloys, which commonly contain a spherically-shaped W matrix with a NiFe-rich binder phase interweaving between matrix particles [69, 86]. While some spherical BCC W particles are observed and attributed to ineffective melting during deposition [87], the FCC NiFeW regions are smaller, more dispersed, and persist both inter- and intra-granularly. The presence of this ductile NiFeW phase eliminated microcracking but produced a high degree of retained porosity.

The third alloy merged the strategy of adding ZrC to refine the microstructure with the introduction of a ductile FCC NiFeW phase in the multicomponent WNiFe+ZrC system. The as-printed alloy exhibited refined grain sizes (≤1 μm) within a two-phase BCC W and FCC NiFeW microstructure but with substantial texturing at the micrometer-scale. This microstructure is reminiscent of a traditionally processed WNiFe heavy alloy, which consists of a NiFe-rich binder region intermixed with elongated W regions [69, 88]. However, the TEM lift-out shown in **Figure 5** was deliberately extracted from a region containing the NiFeW phase, and **Figure 2** shows larger regions of pure W containing W-W grain boundaries exist in the alloy that are not commonly observed in liquid phase sintered WNiFe heavy alloys. Unlike the W+ZrC alloy, where the ZrC additive decomposed to form $ZrO_2$ dispersoids, the addition of ZrC to the WNiFe ternary led to a nearly homogeneous distribution of Zr in the resulting microstructure – albeit with a slight preference for Zr to occupy the BCC W matrix. This behavior is confirmed by comparing the XANES data collected at the Ni K-edge in WNiFe (**Figure 6**a) and WNiFe+ZrC (**Figure 7**a), where the subtle reductions in the edge peak and shifts in the post-edge peaks in the BCC W spectra indicated that Zr occupies substitutional sites in the W BCC solid solution, but not at a high enough concentration to significantly alter the XANES spectra. The refined grain structure, absence of brittle $ZrO_2$ dispersoids, and presence of a ductile NiFeW phase eliminated solidification cracking while reducing the degree of retained porosity relative to the ternary WNiFe alloy.

The Vickers hardness of the AM W+ZrC (460 HV) was comparable to W+ZrC produced through deformation processing routes such as cold-rolled W+(1 wt.%)ZrC (510 HV) [24]. Conversely, lower density sintered W+(0.5-1 wt.%)ZrC exhibited a reduced average hardness in the range of 225-300 HV [21]. The increase in hardness relative to pure W is a reflection of the decreased grain size and the presence of $ZrO_2$ nanodispersoids, which serve to decrease dislocation mobility and consequently increase the hardness through the well-established Orowan strengthening mechanism [89]. Furthermore, the microstructure exhibited brittle behavior through indentation, as expected for the addition of a brittle ceramic to an already brittle metallic material. Annealing the W+ZrC microstructure showed little reduction in the average hardness, decreasing only to ~435 HV at 1350°C (a 6% drop), and consistent with the minimal change in the coherent domain size from XRD. Microstructural stability in this system is a reflection of the reduction in grain boundary motion due to Zener pinning arising from the presence of $ZrO_2$ particles [90]. Furthermore, the addition of ZrC served to decrease the overall thermophysical properties of the composite, reducing the thermal conductivity (**Figure 11**c) at room temperature from nominally 100 W/m×K in pure W to 60 W/m×K. While the addition of a ceramic is expected to reduce thermal conductivity, this value is significantly lower than the cold-rolled W+1.0wt.%ZrC (150 W/m×K) [24] but aligns with the reduction in pure W due to the presence of microstructural heterogeneities inherent to the AM solidification process.

The hardness of 530 HV for the ternary WNiFe alloy was on the upper end of the spectrum exhibited by other typical W heavy alloys from literature, which ranges from 310 HV to 540 HV in the case of liquid phase sintered WNiFeCo [91, 92], and greater than nominally 490 HV as demonstrated by other SLM AM WNiFe microstructures [93]. The presence of the FCC NiFeW phase also produced a more ductile indentation response relative to pure W and the W+ZrC sample, consistent with prior reports on ductile phase-toughened W alloys [94]. Annealing led to a significant reduction in hardness due to relaxation (of approximately 30% in both alloys), as evidenced by the increase in the BCC W lattice parameter, concurrent decrease in the FCC NiFeW lattice parameter, and reduction in the microstrain. In the WNiFe+ZrC system, the introduction of Zr as a substitutional element in both the BCC W and FCC NiFeW phases manifested as a 12% increase in hardness relative to WNiFe. Typically, such an increase in hardness due to the addition of a substitutional element would be attributed primarily to solid solution strengthening; however, prior work on Zr-doped W heavy alloys have suggested that grain refinement is the primary driver of mechanical hardening in these alloys [95, 96].

The EBSD maps in **Figure 3** indeed indicate that the addition of ZrC refined the microstructure, but consistencies in the coherent domain sizes across the different alloys suggest additional mechanisms contribute to the hardening beyond simple grain refinement. The addition of Zr instead stabilized 2D defects as reflected by the increase in microstrain values in **Figure 9**, which accounts for the increased hardness exhibited by the WNiFe+ZrC alloy in both the as-printed and annealed conditions. Directly comparing the hardness of the alloys containing the ductile NiFeW phase with either the pure W or the W+ZrC sample is difficult since microcracking in the latter two samples serve as an energy dissipation mechanism that would otherwise contribute to further plastic deformation, thereby artificially increasing the reported hardness of these samples. Finally, the identical scaling of the thermal conductivity with temperature for these alloys indicates that Zr in solid solution had little effect on thermal conductivity. However, the redistribution of Fe via diffusion from the NiFeW phase at elevated temperature, as inferred from the change in lattice parameter in **Figure 10**, increased the thermal conductivity via a composite effect [97, 98] by reducing the degree of alloying in the NiFeW phase.

## 5. Conclusions

The microstructural and thermophysical impact of two distinct alloying strategies, and their combination, for AM W was explored: ZrC additives to refine the microstructure during solidification, and a ductile FCC NiFe phase to increase toughness and accommodate large thermal strains. While SEM, TEM, and multimodal X-ray characterization demonstrated appreciable variations in the microstructural characteristics of the as-printed samples, in general, the effect of ZrC or NiFe was confirmed to substantially reduce the crack density relative to as-printed commercially pure W, with the following system specific findings:

- ZrC additions to the powder feedstock led to $ZrO_2$ dispersoids, a fine-grained microstructure, and reduced crack density in the solidified part; however, the grain refinement mechanism was not determined in this study. While ZrC additions improved the hardness relative to as-printed commercially pure W, the alloy was brittle, which was confirmed by microcrack formation around the residual indentation impressions.
- The WNiFe-based alloy samples were free of solidification cracks but contained enhanced porosity due to mass loss from selective vaporization of the NiFeW phase during processing. The as-printed hardness was improved due to a combination of solid solution and dispersion strengthening, and the ductile FCC NiFeW phase eliminated microcrack formation around indentation impressions.

Combining these alloying effects in the multicomponent WNiFe+ZrC alloy resulted in a fine-grained microstructure with Zr retained in solid solution rather than forming embrittling oxides. By eliminating the oxide phase and introducing an intrinsically tough FCC NiFeW phase, cracking around the residual impressions was completely suppressed with increased hardness relative to the other alloys. The thermal conductivity of all alloy configurations was reduced relative to AM W, as expected, and with system-dependent temperature scaling behavior that derived from the intrinsic conductivity of the given alloy chemistry or redistribution of solute at elevated temperatures. In summary, both alloying strategies were shown to reduce solidification crack formation with the WNiFe+ZrC alloy balancing favorable attributes from the ZrC grain refiner and NiFe toughening phase as effective pathways for enabling laser AM W materials with enhanced stability and performance.

# 6. Acknowledgements

This work was supported by the U.S. Department of Energy (DOE), Office of Science, Fusion Energy Sciences, under contract DE-SC0017899 with secondary support from contract DE-SC0023096. The authors would like to thank Sarah Dickens for helping with EBSD data collection. This work was performed, in part, at the Center for Integrated Nanotechnologies, an Office of Science User Facility operated for the U.S. Department of Energy (DOE) Office of Science by Los Alamos National Laboratory (Contract 89233218CNA000001) and Sandia

National Laboratories (Contract DE-NA-0003525). This research also used resources in the Electron Microscopy facility of the Center for Functional Nanomaterials, which is a U.S. DOE Office of Science Facility, at Brookhaven National Laboratory under Contract No. DE-SC0012704. The authors also acknowledge use of beamline 3-ID (HXN) and 28-ID-2 (XPD) at NSLS-II, a U.S. DOE Office of Science User Facility operated for the DOE Office of Science by Brookhaven National Laboratory under Contract No. DE-SC0012704. The views expressed in the article do not necessarily represent the views of the U.S. DOE or the United States Government.